\documentclass[14pt]{revtex4}

\usepackage{amsmath}
\usepackage{rotating}
\usepackage{hyperref}

\hypersetup{
	colorlinks   = true, 
	urlcolor     = blue, 
	linkcolor    = blue, 
	citecolor   = cyan 
}

\usepackage{graphicx,epstopdf}

\begin{document}

\title{Does particle creation mechanism favour formation of Black hole or naked singularity? }

\author{Sudipto Bhattacharjee\footnote {\color{blue} slg00sudipto@gmail.com}}

\author{Subhajit Saha\footnote {\color{blue}subhajit1729@gmail.com}}

\author{Subenoy Chakraborty\footnote {\color{blue}schakraborty.math@gmail.com}}

\affiliation{$^{\ast, \ddagger}$ Department of Mathematics, Jadavpur University, Kolkata-700032, West Bengal, India.}
\affiliation{$^{\dagger}$ Department of Physical Sciences,\\ Indian Institute of Science Education and Research Kolkata,\\ Mohanpur 741246, West Bengal, India.}

\begin{abstract}
	
The paper deals with collapse dynamics of a spherically symmetric massive star in the framework of non-equilibrium thermodynamic prescription through particle creation mechanism. The matter content in the star is in the form of perfect fluid  with barotropic equation of state and the dissipative phenomena due to non-equilibrium thermodynamics is in the form of bulk viscosity. For simplicity, the thermodynamic system is chosen to be adiabatic ({\it i.e.,} isentropic) so that the effective bulk viscous  pressure is linearly related to the particle creation rate. As a result, the evolution of the collapsing star also depends on the particle creation rate. By proper choice of creation rate  as a function of the Hubble parameter, it is found that the end state of the collapse may be either a black hole or a naked singularity.\\\\

Keywords: Particle creation, Adiabatic process, Collapsing star, Black hole.

\end{abstract}

\maketitle

\section{INTRODUCTION}

The study of gravitational collapse is an important issue in classical general relativity. Usually, the stellar objects such as white dwarf and neutron star are formed through a collapsing process. Also in astrophysical collapse one should match the interior and exterior space-time of the collapsing object through the proper junction conditions.\par

Long back in 1939 Oppenhiemer and Snyder \cite{os56} initiated the study of gravitational collapse with interior space-time represented by Friedmann like dust solution with a static Schwarzschild exterior. Since then several authors have extended this study of gravitational collapse. In the following, we shall mention some of the important and realistic generalization of the above pioneering  works: (i) Misner and Sharp \cite{misner} considered the perfect fluid collapse with the same static exterior, (ii) using Vaidya's \cite{vaidya} idea of outgoing radiation of the collapsing body, Santos and collaborators \cite{oliveira1, oliveira2, oliveira3, oliveira4} considered dissipative collapsing matter by allowing radial heat flow ({\it i.e., } radiating collapse). On the other hand, Cissoko  et al  \cite{cissoko} and Goncalves \cite{goncalves} studied junction conditions of a non-static collapsing object with a static interior. Gravitational collapse in the presence of dark energy has been investigated by Mota et al \cite{mota} and Cai \cite{cai} et al.\par

The discovery of Hawking radiation has shown a nice interrelationship between black hole and thermodynamics. Subsequently, it is found that there is a deep inner relationship between gravity and thermodynamics. So it is interesting to consider the thermodynamical analysis of a collapsing massive star which sinks under the attraction of its own gravity and at the end of its life cycle either black hole will form or it will appear as a naked singularity, depending on the nature of the initial data. In particular, it is curious to know the validity of the thermodynamical laws during the collapsing astrophysical object.\par

In recent years, a lot of works \cite{sc223, sc212, sc220, sc1503} have been done in cosmology in the perspective of non-equilibrium thermodynamics within the framework of particle creation mechanism. The main motivation of these works is to explain the well known observational evidence that our universe is going through an accelerating phase since recent past. It is found \cite{sc212, sc1503} that by proper choice of the particle creation rate (as function of Hubble parameter) the late time accelerated expansion can be described in the context of Einstein's general relativistic theory (GRT) without introduction of any exotic matter (dark energy). Also recently \cite{sc223} it is shown that the above models not only describe the late phase of the evolution but also describe the entire cosmic evolution since inflation to $\Lambda$CDM model.\par

The present work is also related to the particle creation mechanism but in context of well known astrophysical problem namely the final fate of a massive collapsing star. We shall address the question whether the particle creation mechanism favours formation of BH or helps the collapsing star to become a naked singularity. \par

The plan of the paper is as follows: In Section II, the basic idea of collapsing mechanism has been presented, collapsing solutions and relevant physical properties for various choices of the particle creation rate has been shown in Section III. Section IV deals with junction conditions and relevant physical interpretations with Schwarzschild-de Sitter as the exterior space-time. The thermodynamics of the collapsing star has been discussed in Section V. A field theoretic description has been shown in Section VI. The paper ends with a brief discussion and concluding remarks in Section VII.

\section{The Basic Idea of Collapsing Mechanism} 
In the present work, the matter of the collapsing star is chosen in the form of perfect fluid with barotropic equation of state $p=(\gamma-1)\rho$, while the dissipative phenomena due to non-equilibrium thermodynamics is in the form of bulk viscosity. For simplicity, the thermodynamical system is chosen as isentropic  ({\it i.e.,} adiabatic) in nature so that the entropy per particle is chosen as constant. As a result, the effective bulk viscous pressure is determined by the particle creation rate \cite{sc223, sc212, sc220, sc1503,z61} as
\begin{equation}\label{eq1}
\Pi=-\frac{\Gamma}{3H}(p+\rho),
\end{equation}
where $\Gamma$ is the particle creation rate, $\Pi$ is the effective bulk viscous pressure (due to dissipation), and $H$ is the Hubble Parameter. For simple collapsing situation, we assume the space-time inside the massive core as homogeneous and isotropic {\it i.e.,} the inside geometry is characterized by the flat Friedmann-Robertson-Walker (FRW) model
\begin{equation}\label{eq2}
ds_{-}^2=dt^2-a^2(t)(dr^2+r^2d\Omega_2^2)
\end{equation} 
and it is a particular case of the inhomogeneous Oppenheimer-Snyder model \cite{os56}. Here $a(t)$ is the scale factor and $d\Omega_2^2=d\theta^2+\sin^2\theta{d\phi^2}$ is the metric on unit 2-sphere. Further, in analogy to cosmology where the curvature effects are not important at the early stages of the evolution \cite{llpr}. It is speculated that the same thing happens for the late stages of the collapsing core. The main question that we shall have to address is the end state of collapse --- a black hole (BH) or a naked singularity (NS), {\it i.e.,} the singularity is covered by an apparent horizon or not.

Apparent horizons are space-like surfaces with future point converging null geodesics on both sides of the surface \cite{shgb, pb50}. In fact, the apparent horizon is a trapped surface lying in a boundary of a particular surface $S$. In particular, if $S$ is a two-sphere embedded in a slice $\Sigma$ of space-time $M$, and let $s^\mu$ be the outward-pointing space-like unit normal to $\Sigma$ and $n^{\mu}$, the future pointing time--like unit normal to $\Sigma$ so that $k^{\mu}=s^{\mu}+n^{\mu}$ is a null vector. Then the surfaces will be called marginally trapped surface if $k^{\mu}_{;\mu}=0$ holds everywhere on $S$ \cite{pb50}.

For the present FRW model, the apparent horizon is characterised by \cite{sc47, sc126, pjcl}
\begin{equation} \label{eq3}
R_{,\alpha}R_{,\beta}g^{\alpha\beta}\equiv(r\dot{a})^2-1=0,
\end{equation}
where $R(t,r)=ra(t)$ is the area radius. Further, if the star is assumed to be untrapped initially then the co-moving boundary surface of the star is space-like and we have on $\Sigma$
\begin{equation} \label{eq4}
R_{,\alpha}R_{,\beta}g^{\alpha\beta}\equiv\left\lbrace r_{\Sigma}\dot{a}(t)\right\rbrace ^2-1<0,
\end{equation}
{\it i.e.,} $0<R_iH_i<1$, where $R_i,H_i$ are the initial area radius and Hubble parameter of the collapsing core. Here $r_{\Sigma}$ denotes the boundary of the collapsing star and we have on $\Sigma$:
\begin{equation} \label{eq5}
ds_{\Sigma}^2=d{\tau}^2-R^2(\tau)d{\Omega}_2^2,
\end{equation}
where $\tau=t$ and $R(\tau)=r_{\Sigma}a(\tau)$ is the area radius of the bounding surface. The metric outside the collapsing star in general can be written in the form \cite{cai, cw0503}
\begin{equation} \label{eq6}
ds_+^2=A^2(T,R)dT^2-B^2(T,R)(dR^2+R^2d{\Omega}_2^2).
\end{equation}
\\In view of the exterior space-time, the surface $\Sigma$ can be expressed as $R=R_{0}(T)$. Israel's junction conditions on the boundary have been discussed in details by Cai and Wang \cite{cai, cw0503}. Once dependence of $A$ and $B$ on $T$ and $R$ is known, it is possible to determine the time evolution of $T,R_{0},A$, and $B$ along the hypersurface $\Sigma$.

For gravitational collapse $\dot{a}<0$ and $R(t,r)\equiv ra(t)$ denotes the geometric radius of the two spheres $(t,r)=$constant. The mass function due to Cahill and McVittie \cite{cm11} is defined as
\begin{equation} \label{eq7}
m(r,t)=\frac{R}{2}(1+R_{,\alpha}R_{,\beta}g^{\alpha\beta})=\frac{1}{2}R\dot{R}^2.
\end{equation}
Thus the total mass of the collapsing cloud is
\begin{equation} \label{eq8}
m(\tau)=m(r_{\Sigma},\tau)=\frac{1}{2}R(\tau)\dot{R}^2(\tau)
\end{equation} 
Note that the inequality \ref{eq4} should hold at the initial epoch so that the collapsing process starts from regular initial data. Further, if the above inequality holds throughout the collapsing process then the collapse will evidently not formed BH.

It should be noted that although the total mass given by Eq. \ref{eq8} and the global structure of the BH depends on the space-time geometry outside the star (and also on the matching conditions) but the basic question of BH formation depends crucially on the development of apparent horizon inside the core --- not on the matching conditions and the choice of space-time outside the star. Although in the present work we shall address the question whether a collapsing massive star will become a BH or not at the end stages of its collapse, still we have explicitly mentioned the junction conditions to compare the collapse dynamics for Schwarzschild and Schwarzschild-de Sitter model as the exterior of the collapsing star

Further, it should be mentioned that supermassive BHs (as at the galactic centre) or recently discovered quasar at redshift $z=7.085$ and mass $M=2\times10^9M_{\odot}$ \cite{m474} which is speculated to be formed from huge massive collapsing star cores of population III has extremely large mass due to cosmological accretion mechanism and mergers in the course of their evolution. We only concentrate ourselves to the discussion related to BH and naked singularities, formed from collapsing star cores.

\section{Collapsing solutions}
The basic Friedmann equations for the present model are

\begin{equation} \label{eq9}
3H^2=8{\pi}G{\rho}\quad ~~~~\mbox{and}~~~~ \quad 2\dot{H}=-8{\pi}G(\rho+p+\Pi),
\end{equation}
where the energy-momentum tensor for the matter distribution is
\begin{equation} \label{eq10}
T_{\mu\nu}=(\rho+p+\Pi)u_{\mu}u_{\nu}+(p+\Pi)g_{\mu\nu},
\end{equation}
having conservation equation ({\it i.e.,} ${T_{\nu}^{\mu}}_{;\mu}=0$)
\begin{equation} \label{eq11}
\dot{\rho}+3H(\rho+p+\Pi)=0.
\end{equation}
Here $H=\frac{\dot{a}}{a}<0$ is the Hubble parameter for the collapsing core and $u_{\mu}$ is the unit time-like vector (normalized velocity vector). Now eliminating the dissipative effective from the Friedmann equations \ref{eq9} by using the isentropic condition \ref{eq1}, the collapse dynamics is characterized by the particle creation rate as
\begin{equation} \label{eq12}
\frac{2\dot{H}}{3H^2}=-\gamma\Big(1-\frac{\Gamma}{3H}\Big),
\end{equation}
where equation of state for the perfect fluid namely $p=(\gamma-1)\rho$ has been used (assuming $\gamma\neq0$). In the present work, we  shall choose $\Gamma$ as
\begin{equation} \label{eq13}
\Gamma=\Gamma_3+3\Gamma_0H+\frac{\Gamma_1}{H}
\end{equation}
with $\Gamma_0$, $\Gamma_1$ and $\Gamma_3$ as real constants. This choice of $\Gamma$ is justified from the recent study \cite{sc1503} where it describes the entire evolution of the Universe from inflation to late time acceleration (up to phantom barrier, asymptotically). Using Eq. \ref{eq13} in Eq. \ref{eq12}, the evolution equation for the scale factor becomes
\begin{equation} \label{eq14}
\frac{\ddot{a}}{a}+\Big\{\frac{3\gamma}{2}\Big(1-\Gamma_0\Big)-1\Big\}\frac{{\dot{a}}^2}{a^2}-\frac{\gamma\Gamma_3}{2}\frac{\dot{a}}{a}-\frac{\gamma\Gamma_1}{2}=0
\end{equation}
which on integration gives
\begin{equation} \label{eq15}
H=[-H_0^{-1}+{\mu}\tanh{T}]^{-1},
\end{equation}
and integrating once more, we obtain
\begin{equation} \label{eq16}
\Big(\frac{a}{a_0}\Big)^{\mu\alpha_1}=e^{lT}\bigg[H_0\bigg\{\frac{\Gamma_3}{2\Gamma_1}\cosh{T}-\mu\sinh{T}\bigg\}\bigg]^m.
\end{equation}
In the above solution, we have $\mu^2=\frac{\{12{\Gamma_1}(1-\Gamma_0)+\Gamma_3^2\}}{4\Gamma_1^2}$, $\alpha_{1}=\frac{\gamma\Gamma_1}{2},\quad m=\frac{\mu}{[\mu^2-(\frac{\Gamma_3}{2\Gamma_1})^2]}$, $ l=\frac{H_0^{-1}}{[\mu^2-H_0^{-2}]}$, $T=\mu\alpha_1(t-t_0)$, $H_0=(\frac{\Gamma_3}{2\Gamma_1})^{-1}$, and $a_0$, $t_0$ are constants of integration (with $\Gamma_0\neq1$). The time of collapse $t_c$ when $a=0$ is obtained from the above Eq. \ref{eq16} as
\begin{equation} \label{eq17}
t_c=t_0+\frac{1}{\mu\alpha_1}\Big[\tanh^{-1}\Big(\frac{1}{H_0\mu}\Big)\Big].
\end{equation}
Using this collapsing time the scale factor and Hubble parameter can be respectively written in compact form as
\begin{alignat}{2} \label{eq18}
\bigg(\frac{a}{a_0}\bigg)^{\mu\alpha_1} &=e^{lT}(\cosh{T})^m\bigg[1-\frac{\tanh{T}}{\tanh{T_c}}\bigg]^m \\
\label{eq19}
H &=-H_0\bigg[1-\frac{\tanh{T}}{\tanh{T_c}}\bigg]^{-1}
\end{alignat}.
The negativity of $H$ characterizes the collapsing process under consideration. If $t_{aH}$ is the time of formation of apparent horizon then from Eq. \ref{eq3} the condition for appearance of apparent horizon takes the form
\begin{equation}
R_0H\frac{a}{a_0}=-1, \nonumber
\end{equation}
{\it i.e.,}
\begin{equation} \label{eq20}
R_0H_0e^{(\frac{l}{\mu\alpha_1})T_{aH}}\Big(\cosh{T_{aH}}\Big)^{\big(\frac{m}{\mu\alpha_1}\big)}\bigg[1-\frac{\tanh{T_{aH}}}{\tanh{T_c}}\bigg]^{\big(\frac{m}{\mu\alpha_1}-1\big)}=1
\end{equation}
with $T_{aH}=\mu\alpha_1(t_{aH}-t_0)$. As $\text{tanh}x$ is an increasing function of $x$, so for real solution of the above equation for $t_{aH}$, we must have the following possibilities:
\renewcommand{\theenumi}{\roman{enumi}}%
\begin{enumerate}
	
	\item  $t_c>t_{aH}$ for any real value of $n$    (=$\frac{m}{\mu\alpha_1}-1$)
	
	\item  $t_c<t_{aH}$ or $t_c>t_{aH}$, if n is an even integer
	Note that limiting situation ({\it i.e.,} $t_c=t_{aH}$) is not possible for  the Eq. \ref{eq20}. Thus, depending on the value of $n$, it is possible to have either a BH (i.e $t_c>t_{aH}$) or a NS ({\it i.e.,} $t_c<t_{aH}$).
\end{enumerate}

We shall now discuss the collapse dynamics for the choice $\Gamma_0=1$. In that case, the evolution equation \ref{eq14} simplifies to
\begin{equation} \label{eq21}
\dot{H}=\frac{\gamma}{2}(\Gamma_3H+\Gamma_1)
\end{equation}
which has the solution
\begin{equation} \label{eq22}
\begin{aligned}
H &=-\delta^2+(H_0+\delta^2)e^{-\frac{\gamma\alpha^2}{2}(t-t_0)},
\\
a &=a_0e^{-\delta^2(t-t_0)}\exp\Big[-\frac{2(H_0+\delta^2)}{\gamma\alpha^2}\Big\{e^{-\frac{\gamma\alpha^2}{2}(t-t_0)}-1\Big\}\Big],
\end{aligned}
\end{equation}
where as before $a_0$, $t_0$ are constants of integration, and $\Gamma_3=-\alpha^2$, $\Gamma_1=-\mu^2$, and $\delta^2=\frac{\Gamma_1}{\Gamma_3}$. From the above expression for the scale factor, we see that the present physical process, {\it i.e.,} collapse of a star will take an infinite time for collapse, {\it i.e.,} $t_c=\infty$. Using Eq. \ref{eq3}, the time of formation of apparent horizon is determined from the relation
\begin{equation}
R_0e^{-\delta^2T_{aH}}\Big[\delta^2-(H_0+\delta^2)e^{-\frac{\gamma\alpha^2}{2}T_{aH}}\Big]\exp\Big[-\frac{2(H_0+\delta^2)}{\gamma\alpha^2}\Big\{e^{-\frac{\gamma\alpha^2}{2}T_{aH}}-1\Big\}\Big]=1
\end{equation}
with $T_{aH}=t_{aH}-t_0$. The above equation shows that $t_{aH}$ always has a finite solution and hence the apparent horizon forms much earlier than the time of collapse. So the collapsing process inevitably leads to formation of a BH.

Further, for the present collapsing process the measure of acceleration and the collapsing mass inside the radius '$r$' at time '$t$' are given by 
\begin{alignat}{2}\label{eq24}
\frac{\ddot{a}}{a} &=\{1-\mu^2\alpha_1sech^2{T}\}H^2,  
\\
\label{eq25}
m(r,t) &=\frac{1}{2}R_0^3A^2e^{\big(\frac{3l}{\mu\alpha_1}T\big)}\Big\{\cosh{T}\Big\}^\frac{3m}{\mu\alpha_1}\bigg[1-\frac{\tanh{T}}{\tanh{T}_c}\bigg]^{\big(\frac{3m}{\mu\alpha_1}-2\big)}
\end{alignat}
for $\Gamma_0\neq1$, and

\begin{alignat}{2}\label{eq26}
\frac{\ddot{a}}{a} &= \Big[-\delta^2+(H_0+\delta^2)e^{-\frac{\gamma\alpha^2}{2}(t-t_0)}-\frac{\gamma\alpha^2}{4}\Big]^2-\Big(\frac{\gamma^2\alpha^4}{16}-\frac{\gamma\mu^2}{2}\Big),
\\
\label{eq27}
m(r,t) &= \frac{1}{2}R_0^3e^{-3\delta^2(t-t_0)}\exp\Big[-\frac{6(H_0+\delta^2)}{\gamma\alpha^2}\Big\{e^{-\frac{\gamma\alpha^2}{2}(t-t_0)}-1\Big\}\Big]\Big[-\delta^2+(H_0+\delta^2)e^{-\frac{\gamma\alpha^2}{2}(t-t_0)}\Big]^2
\end{alignat}
when $\Gamma_0=1$. Thus the total mass of the collapsing star at time $\tau$ within the surface $r_{\Sigma}$ is given by $M(\tau)=m(r_{\Sigma},\tau)$. Also from the above expressions for $m(r,t)$, the total collapsed mass inside the apparent horizon can be obtained as


\begin{equation} \label{eq28}
M_{aH}=M(\tau_{aH})=\left\{
\begin{aligned}
&\frac{1}{2H_0}\left[1-\frac{\tanh{T_{aH}}}{\tanh{T_c}}\right] & \mbox{ for } \Gamma_0\neq1 
\\
&-\frac{1}{2\big[-\delta^2+(H_0+\delta^2)e^{-\frac{\gamma \alpha^2}{2}T_{aH}}\big]} & \mbox{ for } \Gamma_0=1.
\end{aligned}
\right.
\end{equation}

We shall now discuss the following particular choices for particle creation rate which are interesting for the present collapse dynamics of a massive star.\\

$\bullet$ $\Gamma=3\Gamma_0H+\frac{\Gamma_1}{H}$: In this case the evolution equation for the scale factor takes the form ($\Gamma_0\neq1$)
\begin{equation} \label{eq29}
\frac{\ddot{a}}{a}+\bigg\{\frac{3\gamma}{2}(1-\Gamma_0)-1\bigg\}H^2-\frac{\gamma\Gamma_1}{2}=0
\end{equation}
whose solution gives the scale factor and the Hubble parameter as
\begin{equation} \label{eq30}
\begin{aligned}
a &= a_0\bigg[1+\frac{3\gamma{H_0}}{2}(1-\Gamma_0)(t-t_0)\bigg]^{\frac{2}{3\gamma(1-\Gamma_0)}} 
\\
H &= \frac{H_0}{\Big[1+\frac{3\gamma H_0}{2}(1-\Gamma_0)(t-t_0)\Big]}
\\
\end{aligned}
\end{equation}
for $\Gamma_0\neq1$, $\Gamma_1=0$, and
\begin{equation} \label{eq31}
\begin{aligned}
a &= a_0\Big[\cosh{\Big\{\frac{3\gamma}{2}\sqrt{\Gamma_2}(1-\Gamma_0)(t-t_0)\Big\}}+\frac{H_0}{\sqrt{\Gamma_2}}\sinh{\Big\{\frac{3\gamma}{2}\sqrt{\Gamma_2}(1-\Gamma_0)(t-t_0)\Big\}}\Big]^{\frac{2}{3\gamma(1-\Gamma_0)}}  
\\
H &= \sqrt{\Gamma_2}\tanh\Big[\tanh^{-1}{\Big(\frac{H_0}{\sqrt\Gamma_2}\Big)}+\frac{3\gamma}{2}\sqrt\Gamma_2(1-\Gamma_0)(t-t_0)\Big],
\\
\end{aligned}
\end{equation}
for $\Gamma_0\neq1$, $\Gamma_1\neq0$
with $\Gamma_2=\frac{\Gamma_1}{3(1-\Gamma_0)}$, $H_1^2=H_0^2-\Gamma_2$. One may note that ,the solution \ref{eq30} can be obtained from the solution \ref{eq31}  in the limiting situation $\Gamma_1\rightarrow0$.

The time of collapse ($t_c$) can be obtained from these solution (by putting $a=0$) as
\begin{equation} \label{eq32}
t_c=\left\{
\begin{aligned}
&t_0-\frac{2}{3\gamma H_0(1-\Gamma_0)}, & \mbox{ for } ~\Gamma_1=0
\\
&t_0-\frac{2}{3\gamma\sqrt{\Gamma_2}(1-\Gamma_0)}\coth^{-1}\Big(\frac{H_0}{\sqrt{\Gamma_2}}\Big), & \mbox{ for }  ~\Gamma_1\neq0. 
\end{aligned}
\right.
\end{equation}
The time of formation of apparent horizon (which is characterized by $RH|_{t=t_{aH}}=-1$) for the present model is given by
\begin{equation} \label{eq33}
t_{aH}=t_0+\frac{2}{3\gamma H_0(1-\Gamma_0)}\Big[-1+\Big(-\frac{1}{R_0H_0}\Big)^{\frac{1}{l}}\Big],~~l={\frac{2}{3\gamma(1-\Gamma_0)}-1}, ~\Gamma_1=0,
\end{equation}
while for $\Gamma_1\neq0$,  $t_{aH}$ is obtained implicitly from the following relation
\begin{eqnarray} \label{eq34}
R_0\sqrt{\Gamma_2}\tanh\Big[\tanh^{-1}{\Big(\frac{H_0}{\sqrt\Gamma_2}\Big)} &+& \frac{3\gamma}{2}\sqrt\Gamma_2(1-\Gamma_0)(t_{aH}-t_0)\Big] \Big[\cosh{\Big\{\frac{3\gamma}{2}\sqrt{\Gamma_2}(1-\Gamma_0)(t_{aH}-t_0)\Big\}} \nonumber \\
&+& \frac{H_0}{\sqrt{\Gamma_2}}\sinh{\Big\{\frac{3\gamma}{2}\sqrt{\Gamma_2}(1-\Gamma_0)(t_{aH}-t_0)\Big\}}\Big]^{\frac{2}{3\gamma(1-\Gamma_0)}}=-1.
\end{eqnarray}

From the above relation it is not possible to obtain an explicit expression for $t_{aH}$. However, in particular, if $\frac{2}{3(1-\Gamma_0)}$ is chosen as unity, then from Eq. \ref{eq34}
\begin{equation} \label{eq35}
t_{aH}=t_0+\frac{1}{\sqrt{\Gamma_2}}\sinh^{-1}\bigg[\frac{\sqrt{\Gamma_2}\pm H_0\sqrt{1-H_1^2R_0^2}}{R_0H_1^2}\bigg]
\end{equation}
provided $H_0<\sqrt{\Gamma_2+\frac{1}{R_0^2}}$. Thus, the time difference between the formation of apparent horizon and the time of collapse (for the choice $\frac{3\gamma(1-\Gamma_0)}{2}=1$)  is given by
\begin{equation} \label{eq36}
t_{aH}-t_c=\left\{
\begin{aligned}
&\frac{1}{H_0}\Big(-\frac{1}{R_0H_0}\Big)^{\frac{1}{l}}, & \mbox{ for } ~\Gamma_1=0
\\ 
&\frac{1}{\sqrt{\Gamma_2}}\sinh^{-1}\bigg[\frac{\sqrt{\Gamma_2}\pm H_0\sqrt{1-H_1^2R_0^2}}{R_0H_1^2}\bigg]+\frac{1}{\sqrt{\Gamma_2}}\coth^{-1}\Big(\frac{H_0}{\sqrt{\Gamma_2}}\Big), & \mbox{ for }~\Gamma_1\neq0 .
\end{aligned}
\right.
\end{equation}\\
Thus, in both the cases no definite conclusion can be made about the nature of collapsing singularity. The acceleration and the mass function have the expressions given by
\begin{equation} \label{eq37}
\frac{\ddot{a}}{a}=\left\{
\begin{aligned}
&\frac{\{1-\frac{3\gamma}{2}(1-\Gamma_0)\}H_0^2}{\Big[1+\frac{3\gamma{H_0}}{2}(1-\Gamma_0)(t-t_0)\Big]^2} \quad & & \mbox{ for } ~\Gamma_1=0
\\
&\Gamma_2 & & \mbox{ for } ~\Gamma_1\neq0\quad(\mbox{ with }~\frac{3\gamma}{2}(1-\Gamma_0)=1),
\end{aligned}
\right.
\end{equation}
and
\begin{equation} \label{eq38}
m(r,t)=\left\{
\begin{split}
&\frac{1}{2}R_0^3H_0^2\Big[1+\frac{3\gamma{H_0}}{2}(1-\Gamma_0)(t-t_0)\Big]^{\{\frac{2}{\gamma(1-\Gamma_0)}-2\}} ~~~\mbox{ for } ~\Gamma_1=0
\\
&\frac{1}{2}R_0^3\Gamma_2\Big[\cosh\big\{\sqrt\Gamma_2(t-t_0)\big\}+\frac{H_0}{\sqrt{\Gamma_2}}\sinh\big\{\sqrt{\Gamma_2}(t-t_0)\big\}\Big]\tanh\big[\tanh^{-1}\left(\frac{H_0}{\sqrt\Gamma_2}\right) \\
& +\sqrt{\Gamma_2}(t-t_0)\big] ~~~\mbox{ for }  ~\Gamma_1\neq0 \mbox{ and }~\frac{3\gamma}{2}(1-\Gamma_0)=1.
\end{split}
\right.
\end{equation}

\begin{figure}
	\centering
	\includegraphics[width=0.63\linewidth]{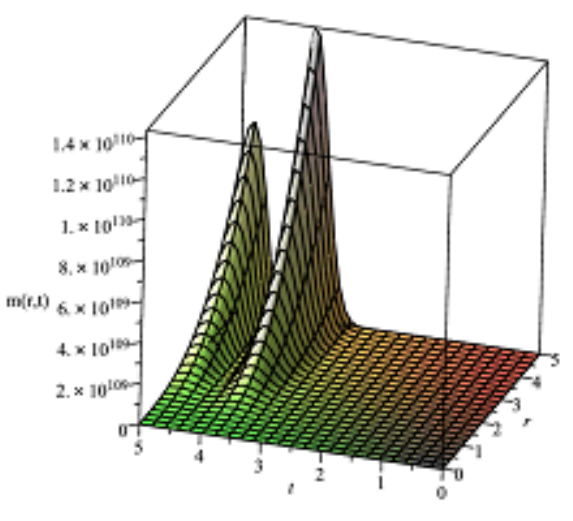}
	\caption{represents $m(r,t)$ (given by \ref{eq27}) against $r$ and $t$ for $\delta=2, \gamma=\frac{4}{3}.$}
\end{figure}
\begin{figure}
	\centering
	\includegraphics[width=0.63\linewidth]{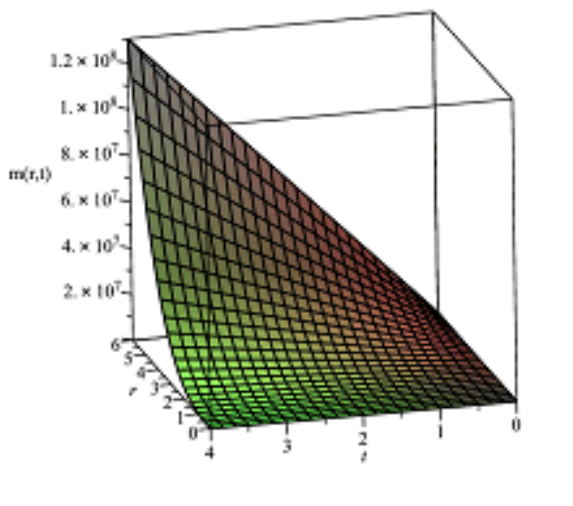}
	\caption{represents $ m(r,t)$ (given by first part of eq .\ref{eq38}) against $r$ and $t$ for $\gamma={4}{3}$ and $\Gamma_0=\frac{2}{3}$.}
\end{figure}

From the above expressions, one can easily estimate the total collapsing mass of the star  having bounding geometric radius $r_\Sigma$ at any time $\tau$ as $M(\tau)=m(r_\Sigma,\tau)$. Also, it is interesting to calculate the mass of the collapsing star inside the apparent horizon as
\begin{equation} \label{eq39}
M_{aH}=M(t_aH)=\left\{
\begin{aligned}
&-\frac{1}{2H_0}\Big[1+\frac{3\gamma H_0}{2}(1-\Gamma_0)(t_{aH}-t_0)\Big] & & \mbox{ for } ~\Gamma_1=0
\\
&-\frac{1}{2\sqrt{\Gamma_2}\tanh\Big[\tanh^{-1}\Big(\frac{H_0}{\sqrt\Gamma_2}\Big)+\frac{3\gamma}{2}\sqrt{\Gamma_2}(1-\Gamma_0)(t_{aH}-t_0)\Big]} & & \mbox{ for } ~\Gamma_1\neq0.
\end{aligned}
\right.
\end{equation}

We shall now consider the choice $\Gamma_0=1$. For this choice, the evolution equation (Eq. \ref{eq29}) simplifies to 
\begin{equation}
\dot{H}=\frac{\gamma\Gamma_1}{2} \nonumber
\end{equation}
which integrating once, we obtain
\begin{equation} \label{eq40}
H=H_0+\frac{\gamma\Gamma_1}{2}(t-t_0).
\end{equation}
As collapsing phase of the star is under consideration, so we choose $\Gamma_1<0$. Hence the scale factor evolves as
\begin{equation} \label{eq41}
a=a_0\exp\Big[H_0(t-t_0)+\frac{\gamma\Gamma_1}{4}(t-t_0)^2\Big].
\end{equation}
It is evident from the above expression of the scale factor that the star requires an infinite time to reach the collapsing singularity ({\it i.e.,} $t_c=\infty$). The time of formation of the apparent horizon ($t_{aH}$) satisfies the relation
\begin{equation} \label{eq42}
R_0\Big[H_0+\frac{\gamma\Gamma_1}{2}T_{aH}\Big]\exp\Big[H_0T_{aH}+\frac{\gamma\Gamma_1}{4}T_{aH}^2\Big]=-1
\end{equation}
for which only finite solution is possible. Thus, in this case, the star will become a BH in a finite time. The expression for acceleration and the mass function are given respectively by 
\begin{equation} \label{eq43}
\frac{\ddot{a}}{a}=\bigg\{H_0+\frac{\gamma\Gamma_1}{2}(t-t_0)\bigg\}^2+\frac{\gamma\Gamma_1}{2}
\end{equation}
and
\begin{equation}\label{eq44}
m(r,t)=\frac{1}{2}R_0^3\exp\Big[3H_0(t-t_0)+\frac{3\gamma\Gamma_1}{4}(t-t_0)^2\Big]\bigg[H_0+\frac{\gamma\Gamma_1}{2}(t-t_0)\bigg]^2.
\end{equation}
\begin{figure}[ht]
	\centering
	\includegraphics[width=0.63\linewidth]{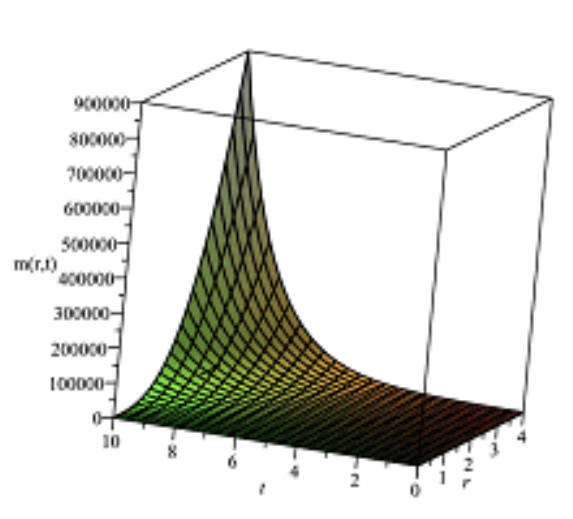}
	\caption{represents $ m(r,t)$ (given by 2nd part of eq. \ref{eq38}) against $r$ and $t$ for $\Gamma_2=0.5,\gamma=\frac{4}{3}$}
\end{figure}

\begin{figure}[ht]
	\centering
	\includegraphics[width=0.63\linewidth]{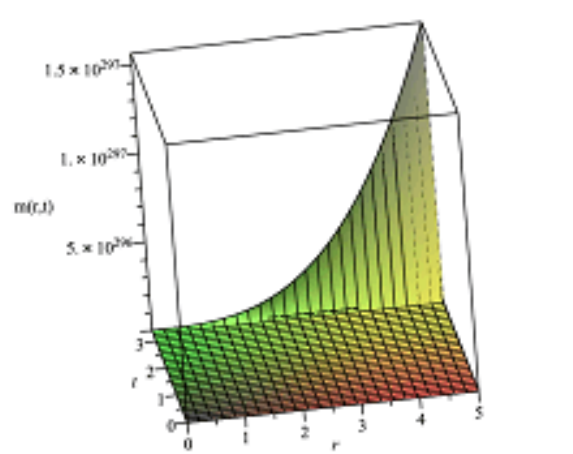}
	\caption{represents the variation of  $ m(r,t)$ (given by eq. \ref{eq44}) against $r$ and $t$ for $\gamma=\frac{4}{3}$ and $\Gamma_1=-1$.}
	
\end{figure}
Hence the mass bounded by the apparent horizon is given by 
\begin{equation} \label{eq45}
M_{aH}=-\frac{1}{2\big[H_0+\frac{\gamma\Gamma_1}{2}T_{aH}\big]},~~T_{aH}=t_{aH}-t_0.
\end{equation}\\

$\bullet$ $\Gamma=3\Gamma_0H+\Gamma_3$: The evolution equation for the scale factor can be obtained from Eq. \ref{eq14} (by putting $\Gamma_1=0$) as 
\begin{equation} \label{eq46}
\frac{\ddot{a}}{a}+\bigg\{\frac{3\gamma}{2}(1-\Gamma_0)-1\bigg\}\frac{\dot{a}^2}{a^2}-\frac{\gamma\Gamma_3}{2}\frac{\dot{a}}{a}=0.
\end{equation}
The relevant physical parameters are given by ($\Gamma_3>0$, $\Gamma_0=1$)
\begin{equation} \label{eq47}
\left.
\begin{aligned}
a &= a_0\exp\Big[\frac{2H_0}{\gamma\Gamma_3}\Big\{e^{\frac{\gamma\Gamma_3}{2}(t-t_0)}-1\Big\}\Big],
\\
H &= H_0\exp\Big[\frac{\gamma\Gamma_3}{2}(t-t_0)\Big],
\\
t_c &= \infty ,
\end{aligned}
\right\}
\end{equation}
and $t_{aH}$ is determined by the relation
\begin{equation} \label{eq48}
R_0H_0e^{\frac{\gamma\Gamma_3}{2}(t_{aH}-t_0)}\exp\bigg[\frac{2H_0}{\gamma\Gamma_3}\bigg\{e^{\frac{\gamma\Gamma_3}{2}(t_{aH}-t_0)}-1\bigg\}\bigg]=-1.
\end{equation}
Also the mass function and the acceleration are respectively given by
\begin{equation} \label{eq49}
\left.
\begin{aligned}
m(r,t) &= \frac{1}{2}R_0^3H_0^2e^{\gamma\Gamma_3(t-t_0)}\exp\bigg[\frac{6H_0}{\gamma\Gamma_3}\bigg\{e^{\frac{\gamma\Gamma_3}{2}(t-t_0)}-1\bigg\}\bigg],
\\
M_{aH} &= -\frac{1}{2H_0}e^{-\frac{\gamma\Gamma3}{2}(t_{aH}-t_0)},
\end{aligned}
\right\}
\end{equation}
and
\begin{equation} \label{eq50}
\frac{\ddot{a}}{a}=H_0e^{\frac{\gamma\Gamma3}{2}(t-t_0)}\big[H_0e^{\frac{\gamma\Gamma3}{2}(t-t_0)}+\frac{\gamma\Gamma_3}{2}\big].
\end{equation}

However, when $\Gamma_0\neq1$, then the expressions for the above parameters are given by ($\Gamma_3>0$, $0<\Gamma_0<1$) 
\begin{equation} \label{eq51}
\left.
\begin{aligned}
a &= a_0\Big[H_0\Big\{\alpha^2e^{\delta(t-t_0)}-\beta^2\Big\}\Big]^\frac{1}{\alpha^2\delta},
\\
H^{-1} &= \alpha^2-\beta^2e^{-\delta(t-t_0)},
\\
t_c &= t_0+\frac{1}{\delta}\ln{\Big(\frac{\beta^2}{\alpha^2}\Big)}.
\end{aligned}
\right\}
\end{equation}
$t_{aH}$ is determined by the relation 
\begin{equation} \label{eq52}
\left.
\begin{aligned}
& R_0(H_0)^{(\frac{1}{\alpha^2\delta})}e^{\delta(t_{aH}-t_0)}\Big[\alpha^2e^{\delta(t_{aH}-t_0)}-\beta^2\Big]^{({\frac{1}{\alpha^2\delta}}-1)}=-1,
\\
m(r,t) &= \frac{1}{2}R_0^3H_0^{(\frac{3}{\alpha^2\delta})}e^{2\delta(t-t_0)}\Big[\alpha^2e^{\delta(t-t_0)}-\beta^2\Big]^{(\frac{3}{\alpha^2\delta}-2)},
\\
M_{aH} &= \frac{1}{2}e^{-\delta(t_{aH}-t_0)}\Big[\beta^2-\alpha^2e^{\delta(t_{aH}-t_0)}\Big],
\\
\frac{\ddot{a}}{a} &= \frac{\Big\{1-\delta\beta^2e^{-\delta(t-t_0)}\Big\}}{\Big\{\alpha^2-\beta^2e^{-\delta(t-t_0)}\Big\}^2}.
\end{aligned}
\right\}
\end{equation}
where $\alpha^2=\frac{3(1-\Gamma_0)}{\Gamma_3},~\beta^2=\alpha^2-\frac{1}{H_0},~\mbox{and} ~\delta=\frac{\gamma\Gamma_3}{2}$.
\begin{figure}
	\centering
	\includegraphics[width=0.63\linewidth]{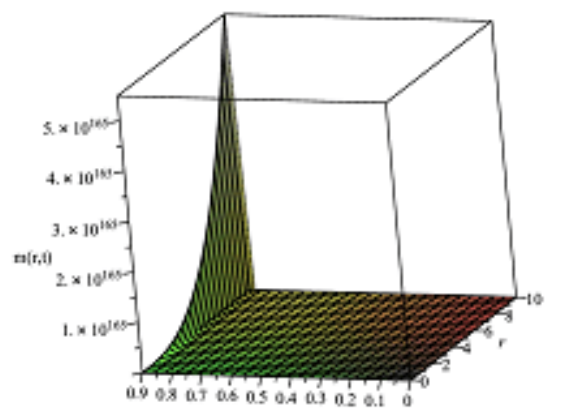}
	\caption{represents  $ m(r,t)$ (given by \ref{eq49}) against $r$ and $t$ for  $\gamma=\frac{4}{3}$ and $\Gamma_3=0.00001$}
\end{figure}
\begin{figure}[!htb]
	\centering
	\includegraphics[width=0.63\linewidth]{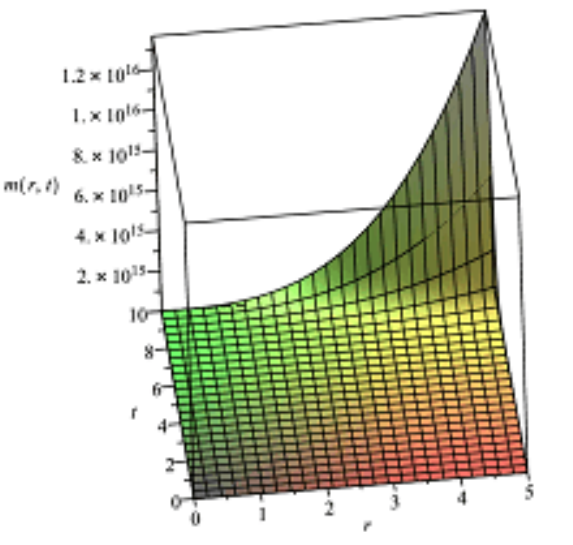}
	\caption{represents  $ m(r,t)$ (given by \ref{eq52}) against $r$ and $t$ for $\Gamma_0=0.5$, $\Gamma_3=1$ and $\gamma=\frac{4}{3}$.}
\end{figure}

Now using the expression for $t_c$ in the determining equation for $t_{aH}$, we have
\begin{equation} \label{eq53}
\frac{R_0}{\alpha^2}\Big(H_0\alpha^2\Big)^{\big(\frac{1}{\alpha^2\delta}\big)}e^{T_{aH}}\Big[e^{T_{aH}}-e^{T_c}\Big]^{(\frac{1}{\alpha^2\delta}-1)}=-1.
\end{equation}
The above equation will be consistent if $n=\frac{\alpha^2\delta}{1-\alpha^2\delta}$ is a +ve integer. For even n we always have $t_{aH}>t_c$ i.e. there will be formation of NS while if n is an odd integer then $t_c<t_{aH}$ i.e. BH will form.

\section{Exterior Schwarzschild-de Sitter Space-time and Junction Conditions at the Surface of the Collapsing star}
Let the surface of the collapsing star be $\Sigma$, a time-like $3D$ hypersurface and it divides the $4D$ space-time into two distinct $4D$ manifolds $M^{\pm}$. The interior of the Collapsing star is the manifold $M^{-}$, described by the flat FRW metric \ref{eq2}, while the exterior manifold $M^{+}$ corresponds to Schwarzschild-de Sitter space-time having line element
\begin{equation} \label{eq54}
{ds_{+}}^2=-U(\rho){dT}^2+\frac{1}{U(\rho)}{d\rho^2}+\rho^2d\Omega_2^2
\end{equation}
with $U(\rho)=1-\frac{2M}{\rho}-\big(\frac{\Lambda}{3}\big){\rho}^2$ where $M$ and $\Lambda$ are constants. Now the intrinsic metric on the surface $\Sigma$ of the collapsing star is chosen as 
\begin{equation} \label{eq55}
{ds_{\Sigma}}^2=-{d\tau}^2+D^2(\tau)d\Omega_2^2.
\end{equation}
According to Santos \cite{san106, san216}, the Israel's junction \cite{is44B} conditions across $\Sigma$ are given by
\renewcommand{\theenumi}{\roman{enumi}}%
\begin{enumerate}
	\item Continuity of the line element across the bounding surface $\Sigma$, {\it i.e.,}
	\begin{equation} \label{eq56}
	(ds_{-}^2)_{\Sigma}=(ds_{+}^2)_{\Sigma}=ds_{\Sigma}^2,
	\end{equation}
	where $(\quad)_{\Sigma}$ means the value of the corresponding quantity on the surface $\Sigma$.
	\item The extrinsic curvature should be continuous through the bounding surface $\Sigma$, {\it i.e.,}
	\begin{equation} \label{eq57}
	K_{ab}^{+}=K_{ab}^{-} \quad \mbox{on} ~~\Sigma .
	\end{equation}
\end{enumerate}
According to Eisenhart \cite{eip},the explicit expression for extrinsic curvature components are 
\begin{equation} \label{eq58}
K_{ab}^{\pm}=-n_{\mu}^{\pm}\frac{\partial^2\chi_{\pm}^{\mu}}{\partial\xi^a\partial\xi^b}-n_{\mu}^{\pm}\Gamma_{\alpha\beta}^{\mu}\frac{\partial\chi_{\pm}^{\alpha}}{\partial\xi^a}\frac{\partial\chi_{\pm}^{\beta}}{\partial\xi^b},
\end{equation} 
where $\chi_{\pm}^{\alpha}$, $\alpha=0,1,2,3$ are the co-ordinates in $M^{\pm}$, $n_{\mu}^{\pm}$ are the components of the normal vector to $\Sigma$ in the co-ordinates $\chi_{\pm}^{\alpha}$, and $\xi^a=(\tau,\theta,\phi)$ are the intrinsic co-ordinates to $\Sigma$.\par 
In view of $M^{-}$, the surface $\Sigma$ is described mathematically as
\begin{equation} \label{eq59}
\Sigma:f_{-}(r,t)=r-r_{\Sigma}=0
\end{equation}
with $r_{\Sigma}$ as constant. As the vector with components $\frac{\partial f}{\partial \chi_{-}^{\alpha}}$ is orthogonal to $\Sigma$, so the unit normal vector is
\begin{equation}
n^{-}_{\alpha}=(0,a,0,0).\nonumber
\end{equation}
Similarly, from the point of view of $M^{+}$, the mathematical description of $\Sigma$ is given by
\begin{equation} \label{eq60}
\Sigma : f_{+}(\rho,T)=\rho-\rho_{\Sigma}(T)=0
\end{equation}
with unit normal vector
\begin{equation} \label{eq61}
n^{+}_{\alpha}=\Big[U-\frac{1}{U}\Big(\frac{d\rho}{dT}\Big)^2\Big]^{-\frac{1}{2}}\Big(-\frac{d\rho}{dT},1,0,0\Big).
\end{equation}
Due to the continuity equation \ref{eq56} we have
\begin{equation} \label{eq62}
\frac{dt}{d\tau}=1,\qquad D(\tau)=R_{\Sigma}={r_{\Sigma}}a
\end{equation}
and
\begin{equation} \label{eq63}
\frac{dT}{d\tau}=\Big\{-U(\rho)+\frac{1}{U(\rho)}\Big(\frac{d\rho}{dT}\Big)^2\Big\}^{-\frac{1}{2}},\quad D(\tau)=\rho_{\Sigma}(T).
\end{equation}
The non-vanishing components of the extrinsic curvature for the interior metric are
\begin{equation} \label{eq64}
K_{\theta\theta}^{-}=\text{cosec}^2\theta K_{\phi\phi}^{-}={r_{\Sigma}}a(t)=R_{\Sigma}
\end{equation}
and those for the exterior space-time on the boundary $\Sigma$ are
\begin{equation} \label{eq65}
K_{\tau\tau}^{+}=\Big[\frac{U\ddot{T}}{\rho}+\frac{dU}{d\rho}\dot{T}\Big]_{\Sigma}
\end{equation}
and
\begin{equation} \label{eq66}
K_{\theta\theta}^{+}=\text{cosec}^2\theta K_{\phi\phi}^{+}=\Big[\dot{T}U\rho\Big]_{\Sigma},
\end{equation}
where an overdot denote differentiation w.r.t. $\tau$. \par
So the junction conditions \ref{eq57} give the following relations (~on $\Sigma$~)
\begin{equation} \label{eq67}
U=1-\dot{R}^2
\end{equation}
and
\begin{equation} \label{eq68}
\dot{T}=\frac{R}{a}(1-R^2H^2)^{-1}.
\end{equation}
Using the explicit expression for $U$, we have from Eq. \ref{eq67},
\begin{equation} \label{eq69}
\frac{1}{2}\dot{R}^2-\frac{M}{R}-\frac{\Lambda}{6}R^2=0\quad \mbox{(on the boundary)}.
\end{equation}
This is nothing but the energy conservation equation on the boundary surface $\Sigma$ \cite{sc}. Note that in the absence of the cosmological term $\Lambda$, {\it i.e.,} if the exterior of the collapsing star is purely Schwarzschild in nature, then from Eq. \ref{eq69}  we see that the Schwarzschild mass $M$ is nothing but the total mass of the collapsing cloud [see Eq. \ref{eq8}] due to Cahill and McVittie \cite{cm11}. Further, the presence of the $\Lambda$-term in Eq. \ref{eq69} shows a repulsive term in the Newtonian potential \cite{bbn23} 
\begin{equation} \label{eq70}
\phi(R)=\frac{M}{R}+\frac{\Lambda}{6}R^2.
\end{equation}
\section{thermodynamics of the Collapsing Star}
Usually, a space-time singularity where physical laws  breakdown ({\it i.e.,} density, curvature and other physical quantities diverge) is not considered in the space-time manifold. So naturally, it is speculated \cite{mpj1106} that the thermodynamical properties on the manifold may not be smooth ({\it i.e.,} regular) in the limit of approaching the singularity. Consequently, one can speculate that the Cosmic Censorship conjecture (CCC) \cite{pen1} may be related to the thermodynamical nature of the space-time manifold near naked singularity \cite{mpj1106}.\par 
For dynamical BHs, Hayward \cite{hay15} introduced the notion of trapping horizon and the idea of Unified First Law. Subsequently, this generalization has been extended to FRW cosmological model showing the equivalence between gravity and thermodynamics at cosmological scenario \cite{cc75}. As trapping horizon coincides with apparent horizon for FRW model so thermodynamical behavior will be investigated at the apparent horizon (formed inside the collapsing sphere) in the limit of approach to the singularity.\par 
We shall now investigate the validity of the second law of thermodynamics in the present context. For BH thermodynamics, the second law states that the area of a future (outer) horizon is non-decreasing provided the energy conditions are satisfied. However, in cosmological context, generalized second law of thermodynamics (GSLT) implies the non-decreasing nature of the total entropy ({\it i.e.,} entropy of the horizon together with the entropy of the matter distribution bounded by the horizon) variation in course of evolution, although there does not have any proper thermodynamical definition of the entropy function considering the microscopic description of space-time. Fortunately, in case of collapse dynamics, one does not have to take into account of the entropy of the in falling matter, only one has to examine the increasing or decreasing nature of the entropy function at the trapping horizon ({\it i.e.,} apparent horizon for FRW model). But if the final fate of the collapsing object is a BH then apparent horizon is an inner trapping horizon and second law of BH thermodynamics can not be applied to it \cite{mpj1106}. To avoid this difficulty one may redefine the trapping horizon of the space-time as the union of the inner apparent horizon and the outer event horizon or in other words as the boundary of the trapped region in the space-time. As irrespective of the final fate of the collapsing object, the volume of the trapped region increases with time, so the entropy of the trapping horizon can be defined as the rate of change of the volume of the trapped region w.r.t. the area radius $R$. Hence when the final fate of a collapsing star is a BH then the entropy of the horizon is defined as \cite{mpj1106}
\begin{equation} \label{eq71}
S_h=\pi(R_{eh}^2-R_{ah}^2),
\end{equation}
where $R_{eh}$ is the Schwarzschild radius of the exterior space-time. On the other hand, in case of naked singularity as the final fate, the horizon entropy is defined as \cite{mpj1106}
\begin{equation} \label{eq72}
S_h=\left\{
\begin{aligned}
&\pi R_{ah}^2 & & \mbox{ for } t\in[t_c,t_{aH})
\\
&\pi R_{eh}^2 & & \mbox{ for } t\in[t_{aH},\infty)
\end{aligned}
\right.
\end{equation}
Now to examine the validity of the second law of thermodynamics we consider the two cases namely that either the exterior space-time is purely Schwarzschild or the Schwarzschild-de Sitter model:\\\\
{\bf a. \quad Schwarzschild space-time outside the collapsing star:}
\\

In this case,
\begin{equation} \label{eq73}
R_{eh}=2M=R\dot{R}^2.
\end{equation}
So,
\begin{equation} \label{eq74}
S_{h}=\pi \left(R^2\dot{R}^4-\frac{1}{H^2}\right)=\pi \left(H^4R^6-\frac{1}{H^2}\right)
\end{equation}
when BH as the end state of collapse, and
\begin{equation} \label{eq75}
S_h=\left\{
\begin{aligned}
&\frac{\pi}{H^2} & & \mbox{ for } t\in[t_c,t_{aH})
\\
&\pi H^4 R^6 & & \mbox{ for } t\in[t_{aH},\infty)
\end{aligned}
\right.
\end{equation}
when collapse leads to a NS. Thus
\begin{equation} \label{eq76}
\dot{S}_h = \left\{
\begin{aligned}
&\pi\bigg[2\bigg(2H^3R^6+\frac{1}{H^3}\bigg)\dot{H}+6H^5R^6\bigg] & & \mbox{ for } t_{aH} < t_c \quad \mbox{ (BH) }
\\
& -\frac{2\pi}{H^3}\dot{H} & & \mbox{ for } t \in [t_c, t_{aH})
\\
&2\pi H^3R^6[2\dot{H}+3H^2] & & \mbox{ for } t \in [t_c, \infty) \quad \mbox{ (NS) }.
\end{aligned}
\right.
\end{equation}
\\

In {\bf Table-I} we have discussed possible final state,  $\dot{S_h}$, restrictions for the validity of the 2nd law of thermodynamics and the Hubble parameter for different particle creation rate in the Schwarzschild space-time outside the collapsing star.

{\bf b. \quad Exterior Schwarzschild-de Sitter space time.}
\\

Using Eq. \ref{eq69}, the Schwarzschild radius of the exterior space-time is given by 
\begin{equation} \label{eq77}
R_{eh}= 2 M_{sd}= R\dot{R}^2-\frac{\Lambda}{3}R^3.
\end{equation}
Hence the entropy of the horizon is given by
\begin{equation} \label{eq78}
S_h =\left\{
\begin{aligned}
& \pi \bigg\{\bigg(H^2-\frac{\Lambda}{3}\bigg)^2R^6-\frac{1}{H^2}\bigg\} & & \mbox{ for } t_{aH} <t_c
\\
& \frac{\pi}{H^2} & & \mbox{ for } t \in [t_c, t_{aH})
\\
& \pi R^6 \bigg(H^2-\frac{\Lambda}{3}\bigg)^2 & & \mbox{ for } t \in [t_{aH},\infty).
\end{aligned}
\right.
\end{equation}
The entropy variation takes the form
\begin{equation} \label{eq79}
\dot{S}_h =\left\{
\begin{aligned}
&\pi \bigg[6HR^6\bigg(H^2-\frac{\Lambda}{3}\bigg)^2+4HR^6\bigg(H^2-\frac{\Lambda}{3}\bigg)\dot{H}+\frac{2}{H^3}\dot{H}\bigg] & & \mbox{ for } t_{aH} < t_c
\\
& -\frac{2\pi}{H^3}\dot{H} & & \mbox{ for } t \in [t_c, t_{aH})
\\
& 2\pi H R^6\bigg(H^2-\frac{\Lambda}{3}\bigg)\bigg[3\bigg(H^2-\frac{\Lambda}{3}\bigg)+2\dot{H}\bigg]  & & \mbox{ for } t \in [t_{aH}, \infty).
\end{aligned}
\right. 
\end{equation}\\
\\
In {\bf Table-II} we have discussed possible final state,  $\dot{S_h}$, restrictions for the validity of the 2nd law of thermodynamics and the Hubble parameter for different particle creation rate in the Exterior Schwarzschild-de Sitter space-time.



\section{a field theoretic description of collapsing process}

The section deals with a description of the collapsing process from the field theoretic point of view, {\it i.e.,} the whole dynamical process, ({\it i.e.,} the collapsing scenario) is considered as the evolution of a scalar field $\phi$ having self interacting potential $V(\phi)$. Equivalently, the collapsing sphere containing effective imperfect fluid can be described by a minimally coupled scalar field. Thus, the energy density and the thermodynamic pressure of the cosmic substrum can be described by the scalar field quantities as
\begin{equation} \label{eq80}
\rho=\frac{1}{2}\dot{\phi}^2+V(\phi) \qquad \mbox{ and} \qquad p_{eff}=p+\Pi=\frac{1}{2}{\dot\phi}^2-V(\phi)
\end{equation}
Hence for the present adiabatic thermodynamical system scalar field quantities can be expressed as
\begin{equation} \label{eq81}
\dot\phi^2=-2\dot{H},
~~~~~~~\mbox{and}
~~~~~~~V(\phi)=3H^2+\dot{H}.
\end{equation}
Note that in the above expression in Eq. \ref{eq81}, we have eliminated the dissipative term $\Pi$ by the isentropic condition equation \ref{eq1}, particle creation rate $\Gamma$ is obtained from the Eq.  \ref{eq13} and we have used the first Friedmann equation \ref{eq9} to eliminate the energy density $\rho$. Now, from the above expression of \ref{eq81} both $\phi$ (integrating first eq. of \ref{eq81})and $V(\phi)$ can be written in parametric form (with $H$ as the parameter) for different particle creation rate as in  {\bf Table-III}.

\section{discussion and Concluding remarks}

The aim of this paper is to analyze the physical process during collapsing phase of a star. The inside matter of the spherical star is chosen as perfect fluid with barotropic equation state $ ~p=(\gamma-1)\rho~~$ ($\gamma\neq0$, a constant). The collapse dynamics is assumed to be a non-equilibrium thermodynamical process having dissipation due to particle creation mechanism. For simplicity, the thermodynamical process is assumed to be isentropic ({\it i.e.,} adiabatic) in nature so that the dissipative effect behaves as bulk viscous pressure and is related linearly to the particle creation rate [see Eq.\ref{eq1}]. Although the choice [in Eq.\ref{eq13}] of the particle creation rate is purely phenomenological but it has some justification from cosmological scenario \cite{sc223, sc1503}. Recently, it has been shown that \cite{sc1503} such choice of particle creation rate can describe the cosmic evolution from inflation to present accelerating phase up to phantom barrier.\par
The curiosity about the collapsing process is due to the lack of definite conclusion about the final fate of the object. It is interesting to examine whether the Cosmic Censorship Conjecture (CCC) of Penrose is obeyed or violated by the collapsing mechanism. We have examined the final fate of the collapse by comparing the time of collapse and the time of formation of the apparent horizon --- whether the final singularity is covered or not by the apparent horizon for various choices of the parameters involved in the choice of the particle creation rate. It is found that in some cases we have definite conclusion about the final phase of collapse (BH or NS) and in other cases depending on some restrictions related to the parameters involved the end state of collapse may be BH or NS.\par
Also we have examined the thermodynamical laws particularly the second law during the collapsing phase. It is found that depending on some restrictions it is possible to have an increase of entropy during the collapse both for BH or NS as the final state of collapse. Interestingly, it is found that in some cases where NS is the definite end state the restrictions in the two time intervals namely $[t_c,t_{aH})$ and $[t_{aH},\infty)$ are contradictory. So the second law of thermodynamics will be violated in any one of the time intervals. But in other cases we do not have such definite conclusion.\par
In the present collapse dynamics of a spherical star, exterior geometry is chosen as Schwarzschild or Schwarzschild-deSitter space-time. The junction conditions on the boundary show a energy conservation equation on it, the corresponding Newtonian force with cosmological constant \cite{ekp314} is
\begin{equation} \label{eq82}
F(R)=-\frac{M}{R^2}+\frac{\Lambda}{3}R.
\end{equation}
As for collapse dynamics, the required force should be attractive in nature, so the area radius $R$ should have an upper bound $\big(\frac{3M}{\Lambda}\big)^{\frac{1}{3}}$. Also, the rate of collapse is given by 
\begin{equation} \label{eq83}
\ddot{R}=-\frac{M}{R^2}+\frac{\Lambda}{3}R.
\end{equation}
From the above expression, we may conclude that the collapsing rate is much faster when the exterior space-time is Schwarzschild rather than the Schwarzschild-de Sitter model. Physically, the cosmological term put some outward pressure to the collapsing star and hence delayed the collapsing process. Also in the context of recent observations the cosmological constant plays the role of dark energy and is best fitted with observations.\par
Finally, one should note that due to the presence of a cosmological constant (DE), a potential barrier is induced into the equation of motion. As a result, particles with a small velocity  will be unable to reach the central object. For future work, it will be interesting to use this idea astrophysically for a particle orbiting a DE black hole so that an estimation of minimum velocity with which the particle can enter the inside of the BH and consequently, the amount of DE inside the BH may be determined.\par


\begin{acknowledgments}
	
	SB is thankful to CSIR, Govt. of India for Junior Research Fellowship. SS is supported by SERB, Govt. of India under National Post-doctoral Fellowship Scheme [File No. PDF/2015/000906]. SC acknowledges UGC-DRS programme at the Department of Mathematics, Jadavpur University. All the authors are thankful to IUCAA, Pune for providing research facilities and warm hospitality while a portion of the work was being done there during a visit.
	
\end{acknowledgments}


	\begin{sidewaystable}[htb]
		\begin{center}
			{\bf Table-I}
		\end{center}
		\tiny
		\begin{center}
			\begin{tabular}{|p{2.8cm}|p{1.4cm}|p{8.4cm}|p{6.5cm}|p{1.6cm}|}
				\hline
				\begin{center}  
					Particle Creation Rate
				\end{center} 
				& 
				\begin{center} 
					Possible final state
				\end{center}
				& 
				\begin{center} 
					$\dot{S_h}$
				\end{center}
				&
				\begin{center} 
					Restrictions for the validity of the 2nd law of thermodynamics 
				\end{center}
				& 
				\begin{center} 
					Hubble Parameter $(H)$
				\end{center}
				\\
				\hline 
				\begin{center}
					$\Gamma=\Gamma_3+3\Gamma_0H+\frac{\Gamma_1}{H}$	
				\end{center}
				&
				\begin{center}
					BH or NS
				\end{center}
				&
				\begin{center} 
					$\pi\bigg[-2\mu^2\alpha_{1}sech^2T\bigg(2H^5R^6+\frac{1}{H}\bigg)+6H^5R^6\bigg], \mbox{ for } t_{aH} < t_c$
				\end{center}
				\begin{center}
					$\frac{2\pi}{H}\mu^2\alpha_{1}sech^2T,  \mbox{ for } t \epsilon [t_c, t_{aH})$
				\end{center}
				\begin{center}
					$ 2\pi H^5R^6[-2\mu^2\alpha_{1}sech^2T+3],  \mbox{ for } t \epsilon [t_c, \infty)$
				\end{center}
				&
				\begin{center}
					$\mu^2\alpha_1 \geq \frac{3H^6R^6}{(2H^6R^6+1)sech^2T},  \mbox{ for } t_{aH} < t_c $
				\end{center}
				\begin{center}
					$\mu^2 \alpha_1 \leq 0,  \mbox{ for } t \epsilon [t_c, t_{aH})$
				\end{center}
				\begin{center}
					$\mu^2 \alpha_1\geq \frac{3}{2sech^2T}, \mbox{ for } t \epsilon [t_c, \infty) $
				\end{center} 
				&
				\begin{center}
					From \ref{eq15}
				\end{center}
				\\
				\hline 
				\begin{center}
					$\Gamma=\Gamma_3+3H+\frac{\Gamma_1}{H}$	
				\end{center}
				&
				\begin{center}
					BH
				\end{center}
				&
				\begin{center} 
					$\pi\bigg[-\gamma\alpha^2\bigg(2H^3R^6+\frac{1}{H^3}\bigg)\bigg(H+\delta^2\bigg)+6H^5R^6\bigg]$
				\end{center}
				&
				\begin{center}
					$\gamma\alpha^2(H+\delta^2) \geq \frac{6H^8R^6}{(2H^6R^6+1)}$
				\end{center} 
				&
				\begin{center}
					From \ref{eq22}
				\end{center}
				\\ 
				\hline
				\begin{center}
					$\Gamma=3\Gamma_0H$		
				\end{center}
				&
				\begin{center}
					BH or NS
				\end{center}
				&
				\begin{center}
					$	3\pi[-\gamma(1-\Gamma_0)(2H^5R^3+\frac{1}{H})+2H^5R^6]$
				\end{center}
				\begin{center} 
					$\frac{3\pi\gamma}{H}(1-\Gamma_0),  \mbox{for} t \epsilon[t_c,t_{aH})$
				\end{center}
				\begin{center}
					$6\pi H^5R^6[-\gamma(1-\Gamma_0)+1], \mbox{for}  t\epsilon[t_c,\infty)$
				\end{center}
				&
				\begin{center}
					$	\gamma(1-\Gamma_0) \geq\frac{6H^6R^6}{2H^6R^3+1}$, for $t_{aH}<t_c$
				\end{center}
				\begin{center}
					$\gamma(1-\Gamma_0) \leq 0,  \mbox{for} t \epsilon[t_c,t_{aH})$
				\end{center}
				\begin{center}
					$\gamma(1-\Gamma_0) \geq 1, \mbox{for}  t\epsilon[t_{aH},\infty)$
				\end{center} 
				&
				\begin{center}
					From \ref{eq30}
				\end{center}
				\\
				\hline 
				\begin{center}
					$\Gamma=3\Gamma_0 H+\Gamma_1/H,$
				\end{center}
				\begin{center}
					$ ~\Gamma_0\neq1,
					~\Gamma_1\neq0,$
				\end{center}
				&
				\begin{center}
					BH or	NS
				\end{center}
				&
				\begin{center}
					$\pi[3\gamma(1-\Gamma_0)(\Gamma_2-H^2)(2H^3R^6+\frac{1}{H^3})+6H^5R^6]$, for $t_{aH}<t_c$
				\end{center}
				\begin{center} 
					$\frac{3\pi\gamma}{H^3}(\Gamma_0-1)(\Gamma_2-H^2),  \mbox{ for } t \epsilon [t_c, t_{aH})$
				\end{center}
				\begin{center}
					$6\pi H^3R^6[\gamma(1-\Gamma_0)(\Gamma_2-H^2)+H^2], \mbox{ for } t \epsilon [t_{aH}, \infty)$
				\end{center}
				&
				\begin{center}
					$\gamma(\Gamma_0-1)(\Gamma_2-H^2)\geq\frac{2H^8R^6}{2H^6R^6+1}$, for $t_{aH}<t_c$
				\end{center}
				\begin{center}
					$\gamma(\Gamma_0-1)(\Gamma_2-H^2) \leq 0, \mbox{ for } t \epsilon [t_c, t_{aH})$
				\end{center}
				\begin{center}
					$\gamma(\Gamma_0-1)(\Gamma_2-H^2)\geq H^2,  \mbox{ for } t \epsilon [t_{aH}, \infty)$
				\end{center} 
				&
				\begin{center}
					From \ref{eq31}
				\end{center}
				\\
				\hline 
				\begin{center}
					$\Gamma=3H+\Gamma_1/H$
				\end{center}
				&
				\begin{center}
					BH
				\end{center}
				&
				\begin{center} 
					$\pi \bigg[\gamma\Gamma_1\bigg(2H^3R^6+\frac{1}{H^3}\bigg)+6H^5R^6\bigg]$
				\end{center}
				&
				\begin{center}
					$\gamma\Gamma_1 \leq -\frac{6H^8R^6}{2H^6R^6+1}$
				\end{center} 
				&
				\begin{center}
					From \ref{eq40}
				\end{center}
				\\
				\hline 
				\begin{center}
					$\Gamma=\Gamma_3+3H,$
				\end{center}
				\begin{center}
					$~\Gamma_3>0$
				\end{center}
				&
				\begin{center}
					BH or NS
				\end{center}
				&
				\begin{center} 
					$\pi \bigg[H\gamma\Gamma_3\bigg(2H^3R^6+\frac{1}{H^3}\bigg)+6H^5R^6\bigg],  \mbox{ for } t_{aH} < t_c$
				\end{center}
				\begin{center}
					$ -\frac{\pi\gamma\Gamma_3}{H^2},  \mbox{ for } t \epsilon [t_c, t_{aH})$
				\end{center}
				\begin{center}
					$ 2\pi H^4R^6[\gamma\Gamma_3+3H],  \mbox{ for } t \epsilon [t_{aH}, \infty)$
				\end{center}
				&
				\begin{center}
					$\gamma\Gamma_3 \geq -\frac{6H^7R^6}{2H^6R^6+1},  \mbox{ for } t_{aH} < t_c$
				\end{center}
				\begin{center}
					$\gamma\Gamma_3 \leq 0, \mbox{ for } t \epsilon [t_c, t_{aH})$
				\end{center}
				\begin{center}
					$\gamma\Gamma_3 \geq -3H ,  \mbox{ for } t \epsilon [t_{aH}, \infty)$
				\end{center} 
				&
				\begin{center}
					From \ref{eq47}
				\end{center}
				\\
				\hline 
				\begin{center}
					$\Gamma=\Gamma_3+3\Gamma_0 H,$
				\end{center}
				\begin{center}
					$ 0<\Gamma_0<1,~\Gamma_3>0,$
				\end{center}
				&
				\begin{center}
					BH or NS
				\end{center}
				&
				\begin{center} 
					$\pi \bigg[2H\delta\bigg(2H^3R^6+\frac{1}{H^3}\bigg)\bigg(1-\alpha^2H\bigg)+6H^5R^6\bigg],  \mbox{ for } t_{aH} < t_c$
				\end{center}
				\begin{center}
					$-\frac{2\pi\delta}{H^2}(1-\alpha^2H),  \mbox{ for } t \epsilon [t_c, t_{aH})$
				\end{center}
				\begin{center}
					$ 2\pi H^4R^6[2\delta+H(3-2\alpha^2\delta)],  \mbox{ for } t \epsilon [t_{aH}, \infty)$
				\end{center}
				&
				\begin{center}
					$(1-\alpha^2H)\delta \geq -\frac{3H^7R^6}{2H^6R^6+1}, \mbox{ for } t_{aH} < t_c$
				\end{center}
				\begin{center}
					$\delta(1-\alpha^2H) \leq 0, \mbox{ for } t \epsilon [t_c, t_{aH})$
				\end{center}
				\begin{center}
					$2\delta+H(3-2\alpha^2\delta)\geq 0,  \mbox{ for } t \epsilon [t_c, \infty)$
				\end{center} 
				&
				\begin{center}
					From \ref{eq51}
				\end{center}
				\\
				\hline
			\end{tabular}
		\end{center}
	\end{sidewaystable}
	\normalsize
	\begin{sidewaystable}[h]
		\begin{center}
			\renewcommand{\arraystretch}{0.5}
			\begin{center}
				{\bf Table-II}
			\end{center}
			\tiny
			\begin{tabular}{|p{2.7cm}|p{1.6cm}|p{8.6cm}|p{8.2cm}|p{1.6cm}|}
				\hline 
				\begin{center}  
					Particle Creation Rate
				\end{center} 
				& 
				\begin{center} 
					Possible final state
				\end{center}
				& 
				\begin{center} 
					$\dot{S_h}$
				\end{center}
				&
				\begin{center} 
					Restrictions for the validity of the 2nd law of thermodynamics 
				\end{center}
				& 
				\begin{center} 
					Hubble Parameter $(H)$
				\end{center}
				\\
				\hline 
				\begin{center}
					$\Gamma=\Gamma_3+3\Gamma_0H+\frac{\Gamma_1}{H}$	
				\end{center}
				&
				\begin{center}
					BH or NS
				\end{center}
				&
				\begin{center} 
					$ \pi\bigg[6HR^6\bigg(H^2-\frac{\Lambda}{3}\bigg)^2-\alpha_{1}\mu^2H^2sech^2T\bigg\{4HR^6\bigg(H^2-\frac{\Lambda}{3}\bigg)+\frac{2}{H^3}\bigg\}\bigg], \mbox{ for } t_{aH} < t_c$
				\end{center}
				\begin{center}
					$\frac{2\pi}{H}\mu^2\alpha_{1}sech^2T, \mbox{ for } t \epsilon [t_c, t_{aH})$
				\end{center}
				\begin{center}
					$ 2\pi H R^6\bigg(H^2-\frac{\Lambda}{3}\bigg)\bigg[3\bigg(H^2-\frac{\Lambda}{3}\bigg)-2\alpha_{1}\mu^2H^2sech^2T\bigg], \mbox{ for } t \epsilon [t_{aH}, \infty)$
				\end{center}
				&
				\begin{center}
					$\mu^2\alpha_1 \geq \frac{3H^2R^6\bigg(H^2-\frac{\Lambda}{3}\bigg)}{\bigg[2H^4R^6\bigg(H^2-\frac{\Lambda}{3}\bigg)+1\bigg]sech^2T},  \mbox{ for } t_{aH} < t_c $
				\end{center}
				\begin{center}
					$\mu^2 \alpha_1 \leq 0, \mbox{ for } t \epsilon [t_c, t_{aH})$
				\end{center}
				\begin{center}
					$\bigg(H^2-\frac{\Lambda}{3}\bigg)\bigg[3(H^2-\frac{\Lambda}{3})-2\alpha_1\mu^2H^2sech^2T\bigg] \leq 0, \mbox{ for } t \epsilon [t_{aH}, \infty) $
				\end{center} 
				&
				\begin{center}
					From \ref{eq15}
				\end{center}
				\\
				\hline 
				\begin{center}
					$\Gamma=\Gamma_3+3H+\frac{\Gamma_1}{H}$	
				\end{center}
				&
				\begin{center}
					BH
				\end{center}
				&
				\begin{center} 
					$\pi\bigg[6HR^6\bigg(H^2-\frac{\Lambda}{3}\bigg)^2-\gamma\alpha^2\bigg(H+\delta^2\bigg)\bigg\{2HR^6\bigg(H^2-\frac{\Lambda}{3}\bigg)+\frac{1}{H^3}\bigg\}\bigg]$
				\end{center}
				&
				\begin{center}
					$6H^4R^6\bigg(H^2-\frac{\Lambda}{3}\bigg)^2-\gamma\alpha^2\bigg(H+\delta^2\bigg)\bigg\{ 2H^4R^6\bigg(H^2-\frac{\Lambda}{3}\bigg)+1\bigg\}  \leq 0$
				\end{center} 
				&
				\begin{center}
					From \ref{eq22}
				\end{center}
				\\ 
				\hline 
				\begin{center}
					$\Gamma=3\Gamma_0H$		
				\end{center}
				&
				\begin{center}
					BH or	NS
				\end{center}
				&
				\begin{center}
					$\pi[6HR^6(H^2-\frac{\Lambda}{3})^2-3\gamma(1-\Gamma_0)\{2H^3R^6(H^2-\frac{\Lambda}{3})+\frac{1}{H}\}]$, for $t_{aH}<t_c$
				\end{center}
				\begin{center} 
					$\frac{3\pi\gamma}{H}(1-\Gamma_0), \mbox{ for } t \epsilon [t_c, t_{aH})$
				\end{center}
				\begin{center}
					$2\pi H R^6\bigg(H^2-\frac{\Lambda}{3}\bigg)\bigg[3\bigg(H^2-\frac{\Lambda}{3}\bigg)-3\gamma\bigg(1-\Gamma_0\bigg)H^2\bigg],  \mbox{ for } t \epsilon [t_{aH}, \infty)$
				\end{center}
				&
				\begin{center}
					$\gamma(1-\Gamma_0)\geq\frac{2H^2R^6(H^2-\frac{\Lambda}{3})^2}{2H^3R^6(H^2-\frac{\Lambda}{3})+1}$, for $t_{aH}<t_c$
				\end{center}
				\begin{center}
					$\gamma(\Gamma_0-1) \geq 0,  \mbox{ for } t \epsilon [t_c, t_{aH})$
				\end{center}
				\begin{center}
					$\bigg(H^2-\frac{\Lambda}{3}\bigg)\bigg[\bigg(H^2-\frac{\Lambda}{3}\bigg)-\gamma\bigg(1-\Gamma_0\bigg)H^2\bigg] \leq 0,  \mbox{ for } t \epsilon [t_{aH}, \infty)$
				\end{center} 
				&
				\begin{center}
					From \ref{eq30}
				\end{center}
				\\
				\hline 
				\begin{center}
					$\Gamma=3\Gamma_0 H+\Gamma_1/H,$
				\end{center}
				\begin{center}
					$ ~\Gamma_0\neq1,
					~\Gamma_1\neq0,$
				\end{center}
				&
				\begin{center}
					BH or NS
				\end{center}
				&
				\begin{center}
					$\pi[6HR^6(H^2-\frac{\Lambda}{3})^2+3\gamma(1-\Gamma_0)(\Gamma_2-H^2)\{2HR^6(H^2-\frac{\Lambda}{3})+\frac{1}{H^3}\}]$, for $t_{aH}<t_c$
				\end{center}
				\begin{center} 
					$ \frac{3\pi\gamma}{H^3}(\Gamma_0-1)(\Gamma_2-H^2),  \mbox{ for } t \epsilon [t_c, t_{aH})$
				\end{center}
				\begin{center}
					$2\pi H R^6\bigg(H^2-\frac{\Lambda}{3}\bigg)\bigg[3\bigg(H^2-\frac{\Lambda}{3}\bigg)+3\gamma\bigg(1-\Gamma_0\bigg)\bigg(\Gamma_2-H^2\bigg)\bigg],  \mbox{ for } t \epsilon [t_{aH}, \infty)$
				\end{center}
				&
				\begin{center}
					$\gamma(1-\Gamma_0)(\Gamma_2-H^2)\leq-\frac{2H^4R^6(H^2-\frac{\Lambda}{3})^2}{2H^4R^6(H^2-\frac{\Lambda}{3})+1}$, for $t_{aH}<t_c$
				\end{center}
				\begin{center}
					$\gamma(\Gamma_0-1)(\Gamma_2-H^2) \leq 0,  \mbox{ for } t \epsilon [t_c, t_{aH})$
				\end{center}
				\begin{center}
					$\bigg(H^2-\frac{\Lambda}{3}\bigg)\bigg[\bigg(H^2-\frac{\Lambda}{3}\bigg)+\gamma\bigg(1-\Gamma_0\bigg)\bigg(\Gamma_2-H^2\bigg)\bigg] \leq 0, \mbox{ for } t \epsilon [t_{aH}, \infty)$
				\end{center} 
				&
				\begin{center}
					From \ref{eq31}
				\end{center}
				\\
				\hline 
				\begin{center}
					$\Gamma=3H+\Gamma_1/H$
				\end{center}
				&
				\begin{center}
					BH
				\end{center}
				&
				\begin{center} 
					$\pi \bigg[6HR^6\bigg(H^2-\frac{\Lambda}{3}\bigg)^2+2\gamma\Gamma_1HR^6\bigg(H^2-\frac{\Lambda}{3}\bigg)+\frac{\gamma\Gamma_1}{H^3}\bigg]$
				\end{center}
				&
				\begin{center}
					$6H^4R^6\bigg(H^2-\frac{\Lambda}{3}\bigg)^2+\gamma\Gamma_1\bigg\{ 2H^4R^6\bigg(H^2-\frac{\Lambda}{3}\bigg)+1\bigg\} \leq 0$
				\end{center} 
				&
				\begin{center}
					From \ref{eq40}
				\end{center}
				\\
				\hline 
				\begin{center}
					$\Gamma=\Gamma_3+3H,$
				\end{center}
				\begin{center}
					$~\Gamma_3>0$
				\end{center}
				&
				\begin{center}
					BH or NS
				\end{center}
				&
				\begin{center} 
					$\pi \bigg[6HR^6\bigg(H^2-\frac{\Lambda}{3}\bigg)^2+2\gamma\Gamma_3H^2R^6\bigg(H^2-\frac{\Lambda}{3}\bigg)+\frac{\gamma\Gamma_3}{H^2}\bigg],  \mbox{ for } t_{aH} < t_c$
				\end{center}
				\begin{center}
					$ -\frac{\pi\gamma\Gamma_3}{H^2},  \mbox{ for } t \epsilon [t_c, t_{aH})$
				\end{center}
				\begin{center}
					$2\pi H R^6\bigg(H^2-\frac{\Lambda}{3}\bigg)\bigg[3\bigg(H^2-\frac{\Lambda}{3}\bigg)+\gamma\Gamma_3H\bigg], \mbox{ for } t \epsilon [t_{aH}, \infty)$
				\end{center}
				&
				\begin{center}
					$6H^3R^6\bigg(H^2-\frac{\Lambda}{3}\bigg)^2+\gamma\Gamma_3\bigg\{ 2H^4R^6\bigg(H^2-\frac{\Lambda}{3}\bigg)+1\bigg\}  \geq 0,  \mbox{ for } t_{aH} < t_c$
				\end{center}
				\begin{center}
					$\gamma\Gamma_3 \leq 0,  \mbox{ for } t \epsilon [t_c, t_{aH})$
				\end{center}
				\begin{center}
					$\bigg(H^2-\frac{\Lambda}{3}\bigg)\bigg[3\bigg(H^2-\frac{\Lambda}{3}\bigg)+\gamma\Gamma_3H\bigg] \leq 0, \mbox{ for } t \epsilon [t_{aH}, \infty)$
				\end{center} 
				&
				\begin{center}
					From \ref{eq47}
				\end{center}
				\\
				\hline 
				\begin{center}
					$\Gamma=\Gamma_3+3\Gamma_0 H,$
				\end{center}
				\begin{center}
					$ 0<\Gamma_0<1,~\Gamma_3>0,$
				\end{center}
				&
				\begin{center}
					BH or NS
				\end{center}
				&
				\begin{center} 
					$\pi \bigg[6HR^6\bigg(H^2-\frac{\Lambda}{3}\bigg)^2+4\delta H^2R^6\bigg(H^2-\frac{\Lambda}{3}\bigg)\bigg(1-\alpha^2H\bigg)+\frac{2\delta}{H^2}\bigg(1-\alpha^2H\bigg)\bigg],  \mbox{ for } t_{aH} < t_c$
				\end{center}
				\begin{center}
					$-\frac{2\pi\delta}{H^2}(1-\alpha^2H),  \mbox{ for } t \epsilon [t_c, t_{aH})$
				\end{center}
				\begin{center}
					$ 2\pi H R^6\bigg(H^2-\frac{\Lambda}{3}\bigg)\bigg[3\bigg(H^2-\frac{\Lambda}{3}\bigg)+2\delta H\bigg(1-\alpha^2H\bigg)\bigg],  \mbox{ for } t \epsilon [t_{aH}, \infty)$
				\end{center}
				&
				\begin{center}
					$3H^3R^6\bigg(H^2-\frac{\Lambda}{3}\bigg)^2+2\delta H^4R^6\bigg(H^2-\frac{\Lambda}{3}\bigg)\bigg(1-\alpha^2H\bigg)+\delta\bigg(1-\alpha^2H\bigg) \geq 0,  \mbox{ for } t_{aH} < t_c$
				\end{center}
				\begin{center}
					$\delta(1-\alpha^2H) \leq 0,  \mbox{ for } t \epsilon [t_c, t_{aH})$
				\end{center}
				\begin{center}
					$\bigg(H^2-\frac{\Lambda}{3}\bigg)\bigg[3\bigg(H^2-\frac{\Lambda}{3}\bigg)+2\delta H\bigg(1-\alpha^2H\bigg)\bigg] \leq 0, \mbox{ for } t \epsilon [t_{aH}, \infty)$
				\end{center} 
				&
				\begin{center}
					From \ref{eq51}
				\end{center}
				\\
				\hline
			\end{tabular}
		\end{center}
	\end{sidewaystable}
	\normalsize
	\begin{table}[h]
		\begin{center}
			{\bf Table-III}
		\end{center}
		\scriptsize
		\begin{tabular}{|p{3.1cm}|p{6.1cm}|p{5cm}|p{1.6cm}|}
			\hline 
			\begin{center}  
				Particle Creation Rate
			\end{center} 
			& 
			\begin{center} 
				Scalar field
			\end{center} 
			\begin{center} 
				($\phi$) 
			\end{center}
			&
			\begin{center} 
				Self interacting potential 
			\end{center}
			\begin{center}
				($V(\phi)$) 
			\end{center}
			& 
			\begin{center} 
				Hubble Parameter $(H)$
			\end{center}
			\\
			\hline 
			\begin{center}
				$\Gamma=\Gamma_3+3\Gamma_0H+\frac{\Gamma_1}{H}$	
			\end{center}
			&
			\begin{center} 
				$\frac{2}{\sqrt{3{\gamma}(1-\Gamma_0)}}\cosh^{-1}\Big[\frac{1}{\mu\Gamma_2}\Big\{H-\frac{\Gamma_3}{6(1-\Gamma_0)}\Big\}\Big]$
			\end{center}
			&
			\begin{center}
				$3H^2\big[(1-\frac{\gamma}{2})+\frac{\gamma\Gamma_0}{2}\big]+\frac{\gamma\Gamma_3}{2}H+\frac{\gamma\Gamma_1}{2}$
			\end{center} 
			&
			\begin{center}
				From \ref{eq15}
			\end{center}
			\\
			\hline 
			\begin{center}
				$\Gamma=\Gamma_3+3H+\frac{\Gamma_1}{H}$	
			\end{center}
			&
			\begin{center} 
				$\frac{4}{\gamma\Gamma_3}\sqrt{-\gamma\Gamma_3H-\gamma\Gamma_1}$
			\end{center}
			&
			\begin{center}
				$3H^2+\frac{\gamma\Gamma_3}{2}H+\frac{\gamma\Gamma_1}{2}$
			\end{center} 
			&
			\begin{center}
				From \ref{eq22}
			\end{center}
			\\
			\hline 
			\begin{center}
				$\Gamma=3\Gamma_0H$		
			\end{center}
			&
			\begin{center} 
				$-\frac{2}{\sqrt{3\gamma(1-\Gamma_0)}}\ln{|H|}$
			\end{center}
			&
			\begin{center}
				$3H^2\big[(1-\frac{\gamma}{2})+\frac{\gamma\Gamma_0}{2}\big]$
			\end{center} 
			&
			\begin{center}
				From \ref{eq30}
			\end{center}
			\\
			\hline 
			\begin{center}
				$\Gamma=3\Gamma_0 H+\Gamma_1/H,$
			\end{center}
			\begin{center}
				$ ~\Gamma_0\neq1,
				~\Gamma_1\neq0,$
			\end{center}
			&
			\begin{center} 
				$\frac{2}{\sqrt{3\gamma(1-\Gamma_0)}}\cosh^{-1}\Big[\sqrt{\frac{3(1-\Gamma_0)}{\Gamma_1}}H\Big]$
			\end{center}
			&
			\begin{center}
				$3H^2\big[(1-\frac{\gamma}{2})+\frac{\gamma\Gamma_0}{2}\big]+\frac{\gamma\Gamma_1}{2}$
			\end{center} 
			&
			\begin{center}
				From \ref{eq31}
			\end{center}
			\\
			\hline 
			\begin{center}
				$\Gamma=3H+\Gamma_1/H$
			\end{center}
			&
			\begin{center} 
				$\frac{2H}{\sqrt{-\gamma\Gamma_1}}$
			\end{center}
			&
			\begin{center}
				$3H^2+\frac{\gamma\Gamma_1}{2}$
			\end{center} 
			&
			\begin{center}
				From \ref{eq40}
			\end{center}
			\\
			\hline 
			\begin{center}
				$\Gamma=\Gamma_3+3H,$
			\end{center}
			\begin{center}
				$~\Gamma_3>0$
			\end{center}
			&
			\begin{center} 
				$4\sqrt{-\frac{H}{\gamma\Gamma_3}}$
			\end{center}
			&
			\begin{center}
				$3H^2+\frac{\gamma\Gamma_3}{2}H$
			\end{center} 
			&
			\begin{center}
				From \ref{eq47}
			\end{center}
			\\
			\hline 
			\begin{center}
				$\Gamma=\Gamma_3+3\Gamma_0 H,$
			\end{center}
			\begin{center}
				$ 0<\Gamma_0<1,~\Gamma_3>0,$
			\end{center}
			&
			\begin{center} 
				$-\frac{2}{\sqrt{3\gamma(1-\Gamma_0)}}\cosh^{-1}\Big[\frac{6(1-\Gamma_0)}{\Gamma_3}H-1\Big]$
			\end{center}
			&
			\begin{center}
				$3H^2\big[(1-\frac{\gamma}{2})+\frac{\gamma\Gamma_0}{2}\big]+\frac{\gamma\Gamma_3}{2}H$
			\end{center} 
			&
			\begin{center}
				From \ref{eq51}
			\end{center}
			\\
			\hline
		\end{tabular}
	\end{table}
	
	\normalsize
\end{document}